\documentclass[prl,twocolumn,groupedaddress,showpacs]{revtex4}

\usepackage{amsmath}
\usepackage{amssymb}
\usepackage{amsfonts}
\usepackage{mathrsfs}
\usepackage{epsfig}
\usepackage{bm}

\begin{document}

\title{Magnetically assisted self-injection and radiation generation for plasma based acceleration}

\author{J. Vieira$^1$}
\email{jorge.vieira@ist.utl.pt}
\author{J. L. Martins$^1$}
\author{V.B. Pathak$^1$}
\author{R. A. Fonseca$^{1,2}$}
\author{W.B. Mori$^3$}
\author{L. O. Silva$^1$}
\email{luis.silva@ist.utl.pt}

\affiliation{$^1$GoLP/Instituto de Plasmas e Fus\~{a}o Nuclear-Laborat\'orio Associado,  Instituto Superior T\'{e}cnico, Lisboa, Portugal}
\affiliation{$^2$DCTI/ISCTE Lisbon University Institute, 1649-026 Lisbon, Portugal}
\affiliation{$^3$Department of Physics and Astronomy, UCLA, Los Angeles, California 90095, USA}


\today

\begin{abstract}
It is shown through analytical modeling and numerical simulations that external magnetic fields can relax the self-trapping thresholds in plasma based accelerators. In addition, the transverse location where self-trapping occurs can be selected by adequate choice of the spatial profile of the external magnetic field. We also find that magnetic-field assisted self-injection can lead to the emission of betatron radiation at well defined frequencies. This controlled injection technique could be explored using state-of-the-art magnetic fields in current/next generation plasma/laser wakefield accelerator experiments.
\end{abstract}

\maketitle

\section{\label{sec:introduction}Introduction}

Plasma based accelerators (PBA) use high intensity laser pulses~\cite{bib:tajima_prl_1979}, with intensities above $I\sim10^{18}~\mathrm{W/cm}^2$, or highly charged particle bunch drivers~\cite{bib:chen_prl_1986}, with more than $10^{10}$ charged particles, to excite ultra-relativistic plasma waves. The ideal plasma density %
to maximize charge and energy gain 
depends on the nature of the driver (i.e. lepton, hadron or laser pulse), typically ranging between $n_0=10^{14}-10^{19}~\mathrm{cm}^{-3}$. At these plasma densities, charged particle bunches can be accelerated by plasma wakefields to 1-100~GeV in 1-100~cm~\cite{bib:lu_prstab_2007}.

The proof-of-principle of PBA is firmly demonstrated~\cite{bib:leemans_natphys_2006,bib:blumenfeld_nature_2007}. %
Presently, 
connection with applications~\cite{bib:patel_nature_2007} is an essential step to further improve this technology. To this end, fine control over the properties of the accelerated electrons is required. Several techniques 
were
proposed to control self-trapping. %
Control over charge and energy of accelerated bunches can be reached using
plasma ramps~\cite{bib:geddes_prl_2008}, counter propagating lasers \cite{bib:faure_nature_2006}, to ionization mechanisms~\cite{bib:pak_prl_2010}, and resorting to non-linear optical effects such as self-focusing~\cite{bib:kalmykov_prl_2009}.

A novel technique using transverse magnetic fields to relax self-injection thresholds has been recently proposed~\cite{bib:vieira_prl_2011}. The use of external magnetic fields in plasma acceleration was first proposed to extend the acceleration distances in plasma accelerators in the surfatron model~\cite{bib:katsouleas_prl_1983}. The role of external magnetic fields in PBA was also explored in~\cite{bib:ren_pop_2004}, and the use of longitudinal magnetic fields to enhance the self-injected charge in laser wakefield acceleration was investigated in~\cite{bib:hosokai_prl_2006}.

This paper presents a detailed derivation of the self-trapping threshold condition in the presence of external fields. Using the particle-in-cell (PIC) code Osiris~\cite{bib:fonseca_book}, it is shown that magnetic injection can be used to generate single or multiple off-axis self-injected bunches with well defined radial injection positions. Using the post-processing radiation code JRad~\cite{bib:martins_spie_2009} it is demonstrated that these electrons may emit betatron radiation at well defined frequencies close to the undulator regime. This paper is structured as follows. In Sec.~\ref{sec:review} describes an analytical trapping condition in the presence of external fields. In Sec.~\ref{sec:simulations}, 3D PIC simulation results are employed to analyze the relevant physical mechanisms of magnetic field assisted self-injection. The use of different B-field geometries to control transverse features of magnetically injected electrons is described in Sec.~\ref{sec:tailor}. Section~\ref{sec:radiation} shows that magnetically assisted injection can lead to the emission of clearly defined betatron radiation harmonics for the first time in PBAs. Finally, conclusions are stated in Sec.~\ref{sec:conclusions}.

\section{\label{sec:review}Trapping conditions in the presence of external fields}

The dynamics of the electrons in the fields created by an intense laser in the blowout regime can be described using Hamiltonian dynamics~\cite{bib:pak_prl_2010}. A general trapping condition in the presence of external fields can be found by examining the evolution of the Hamiltonian of plasma electrons in the co-moving frame, $(x=x,y=y,\xi=v_{\phi}t-z,s=z)$, given by $\mathcal{H}=H-v_{\phi} P_{\|}$, where $(x,y)$ are the transverse coordinates, $z$ the distance, $v_{\phi}$ the wake phase velocity (determined by the driver group velocity), $P_{\|}$ the longitudinal canonical momentum, $H = \sqrt{m_e^2 c^4+(\mathbf{P}+e \mathbf{A})^2}-e \phi$ the Hamiltonian of a charged particle in the presence of electric and magnetic fields, $m_e$ and $e$ the electron mass and charge, $c$ the speed of light, $\mathbf{A}$ and $\phi$ the vector and scalar potentials, and $\mathbf{P}=\mathbf{p}-e \mathbf{A}$, where $\mathbf{P}$ and $\mathbf{p}$ are the canonical and linear momentum respectively.
Normalized units will be used henceforth unless explicitly stated. Mass and charges are normalized to $m_e$ and $e$, respectively, velocity $v$ to $c$, time to $\omega_p=\sqrt{4 \pi n_0 e^2/m_e}$, momentum to $m_e c$, and density to the background plasma density $n_0$. Vector and scalar potentials are normalized to $e/m_e c^2$ and $e/m_e c$ respectively. Magnetic fields (B) are normalized to $\omega_c/\omega_p$ where $\omega_c=e B /m_e$ is the cyclotron frequency.

In order to derive a trapping threshold condition in the presence of external fields, we consider first the expression for the temporal evolution of $\mathcal{H}$:
\begin{equation}
\label{eq:hamilton_1}
\frac{\mathrm{d}\mathcal{H}}{\mathrm{d} t} = (v_{\phi}-v_{\|}) \frac{\mathrm{d}\mathcal{H}}{\mathrm{d}\xi}=\frac{\partial H}{\partial s } = \left[\mathbf{v}\cdot \frac{\partial \mathbf{A}}{\partial s}-\frac{\partial \phi}{\partial s}\right],
\end{equation}
Integration over the particle trajectory yields:
\begin{equation}
\label{eq:hamilton_2}
\mathcal{H}_f-\mathcal{H}_i = \int \mathrm{d} t \frac{\mathrm{d}\mathcal{H}}{\mathrm{d} t} = \int \frac{\mathrm{d} \xi}{v_{\phi}-v_{\|}} \frac{\mathrm{d} \mathcal{H}}{\mathrm{d} t},
\end{equation}
where the subscripts '$i$` and '$f$` refer to the initial and final (trapped) electron positions. The integration is performed along the electron trajectory, and $\mathrm{d} t = \mathrm{d} \xi/(v_{\phi}-v_z)$. Combining Eq.~(\ref{eq:hamilton_1}) with Eq.~(\ref{eq:hamilton_2}) gives:
\begin{equation}
\label{eq:hamilton_3}
\mathcal{H}_f-\mathcal{H}_i = \int \mathrm{d} \xi \frac{\mathrm{d}\mathcal{H}}{\mathrm{d} \xi} = \int \mathrm{d} \xi\left[\mathbf{v}\cdot \frac{\partial \mathbf{A}}{\partial s}-\frac{\partial \phi}{\partial s}\right],
\end{equation}
Using the definition for $\mathcal{H}$, and considering that initially electrons are at rest (i.e. $\mathbf{p}_i = 0$):
\begin{equation}
\label{eq:hamilton_4}
\mathcal{H}_f-\mathcal{H}_i  =  \gamma_{f} -v_{\phi} p_{f\|} - 1 - \left(\phi_f  - v_{\phi} A_{f\|} \right) - \left(\phi_i  - v_{\phi} A_{i\|} \right),
\end{equation}
with $\phi=\phi^{\mathrm{pl}}+\phi^{\mathrm{ext}}$, $A_{\|}=A_{\|}^{\mathrm{pl}}+A_{\|}^{\mathrm{ext}}$, and where the superscripts '$\mathrm{pl}$` and '$\mathrm{ext}$` refer to the plasma and external fields respectively, and $\gamma=(1-\mathbf{v}^2)^{-1/2}$ is the relativistic factor. Defining the wake potential $\psi=\phi-v_{\phi} A_{\|}$, $\Delta \psi = \psi_f-\psi_i$, and assuming that for trapping  the longitudinal velocity of the electron must reach the velocity of the wake i.e. $v_{z}=v_{\phi}$, Eq.~(\ref{eq:hamilton_4}) readily becomes:
\begin{equation}
\label{eq:hamilton_5}
\mathcal{H}_f-\mathcal{H}_i  =  \frac{\gamma}{\gamma_{\phi}^2} - 1 - \Delta \psi^{\mathrm{pl}} -\Delta \psi^{\mathrm{ext}}.
\end{equation}
where $\gamma_{\phi}=(1-v_{\phi}^2)^{-1/2}$ is the gamma factor of the plasma wave. Using Eq.~(\ref{eq:hamilton_3}) to express $\mathcal{H}_f-\mathcal{H}_i$ leads to the trapping condition~\cite{bib:vieira_prl_2011}:
\begin{equation}
\label{eq:btrapping}
1+\Delta \psi^{\mathrm{pl}} = \frac{\gamma}{\gamma_{\phi}^2}-\int\frac{\mathrm{d}\mathcal{H}}{\mathrm{d}\xi}\mathrm{d}\xi-\Delta \psi^{\mathrm{ext}}.
\end{equation}
Equation~(\ref{eq:btrapping}) is a general trapping condition in the presence of external fields, and valid beyond the range of validity of the quasi-static approximation~\cite{bib:sprangle_prl_1990}. Analytical solutions to Eq.~(\ref{eq:btrapping}), however, are not yet known because the calculation of $\Delta \psi^{\mathrm{pl}}$ and $\int\mathrm{d}_{\xi}\mathcal{H} \mathrm{d}{\xi}$ requires accurate prediction of the particle trajectories and field structures at the back of the bubble, where the applicability of the standard analytical models~\cite{bib:lu_prl_2006} is limited. 

To retrieve a general trapping threshold in the absence of external fields, and in the conditions where the quasi-static approximation is valid, it should be considered $\Delta \psi^{\mathrm{ext}}=0$, and $\int \mathrm{d}_{\xi}\mathcal{H}\mathrm{d}\xi=0$ in Eq.~(\ref{eq:btrapping}). For an ultra-relativistic plasma wave with $\gamma_{\phi}\rightarrow\infty$ trapping occurs when $1+\Delta \psi^{\mathrm{pl}}=0$, or, equivalently, $\Delta \psi^{\mathrm{pl}} = -1$. Generally, this condition can only be met at the back of the plasma wave, in regions of maximum accelerating fields, where $\psi^{\mathrm{pl}}$ is minimum and approaches $\psi^{\mathrm{pl}}=-1$~\cite{bib:pak_prl_2010}. In the presence of static external fields the trapping condition becomes $1+\Delta \psi^{\mathrm{pl}}=-\Delta \psi^{\mathrm{ext}}$. The trapping thresholds are relaxed because they can be met when $\Delta \psi^{\mathrm{pl}}$ is larger than $-1$ provided that $\Delta \psi^{\mathrm{ext}}>0$. In other words, trapping may occur for lower values of peak accelerating gradients. Moreover, trapping may be suppressed if $\Delta \psi^{\mathrm{ext}}<0$.


Trapping can also be relaxed (or suppressed) when the external fields vary spatially in $z$ because of the contribution of finite $\int \mathrm{d}_{\xi}\mathcal{H}\mathrm{d}\xi\ne 0$ to Eq.~(\ref{eq:btrapping}). If the profile of the external fields profile leads to $\int \mathrm{d}_{\xi}\mathcal{H}\mathrm{d}\xi>0$ ($\int \mathrm{d}_{\xi}\mathcal{H}\mathrm{d}\xi<0$) along the electron trajectory then trapping is facilitated (suppressed)~\cite{bib:kalmykov_prl_2009,bib:vieira_prl_2011}. Physically, the fact that $\int \mathrm{d}_{\xi}\mathcal{H}\mathrm{d}\xi>0$ is typically associated with the reduction of the wake phase velocity through the accordion effect, thus facilitating self-injection~\cite{bib:kalmykov_prl_2009,bib:vieira_prl_2011,bib:katsouleas_pra_1986}.


\section{\label{sec:simulations}Magnetically controlled self-injection in LWFAs and PWFAs}

\begin{figure}[htbp]
\begin{center}
\includegraphics[width=\columnwidth]{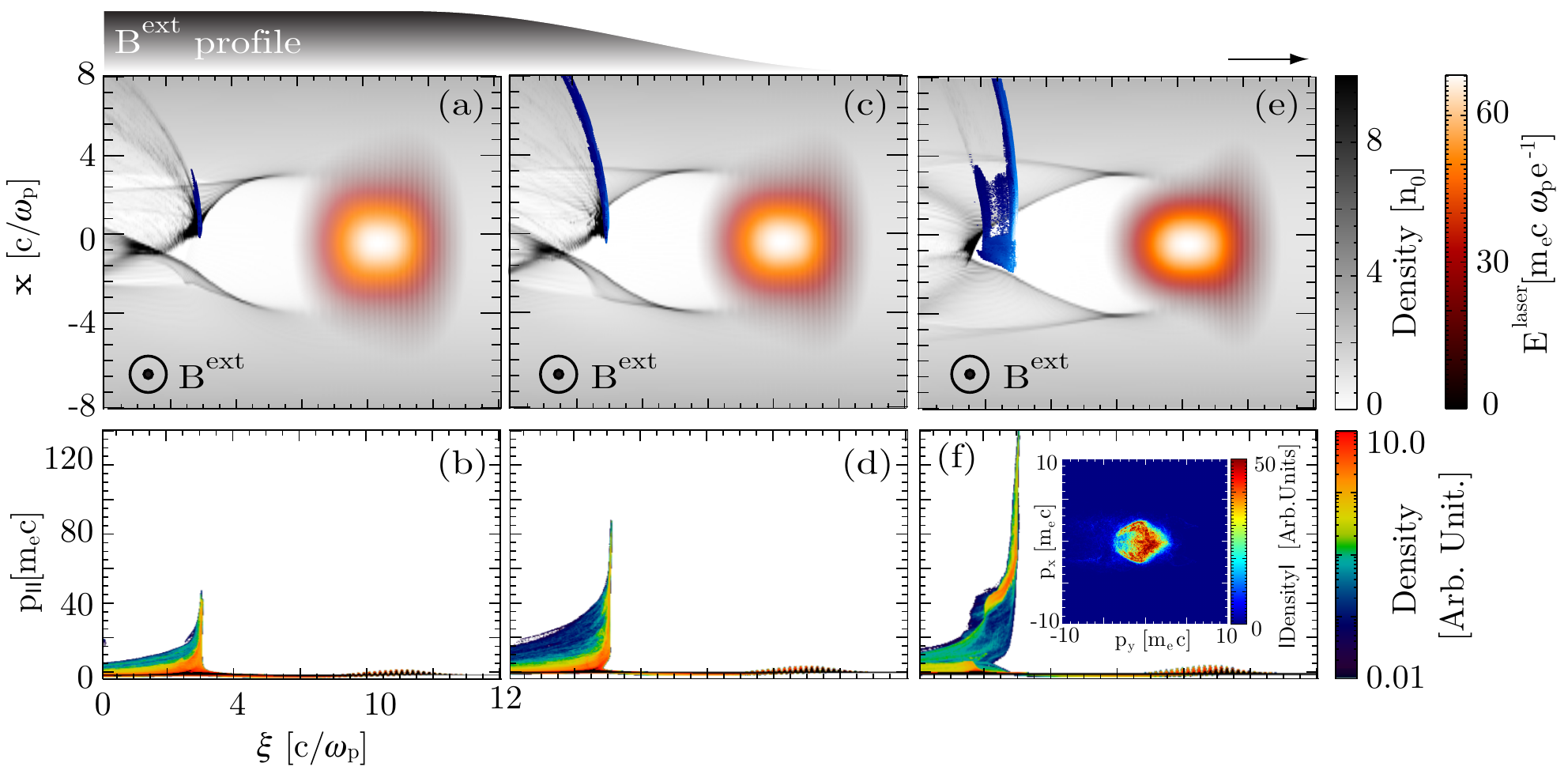}
\caption{\label{fig:lwfa} 3D osiris simulation results illustrating the magnetic self-injection mechanisms. (a), (c), and (e) show the electron plasma density in gray, the self-trapped particles in blue, and the laser pulse envelope in red colors at $t=110/\omega_p$, $t=126 /\omega_p$ and $t=159 /\omega_p$. (b), (d), and (f) show the corresponding $p_{\|}-\xi$ phase-space. The magnetic field leads to off-axis self-injection in a narrow angular region. The inset in (f) represents the transverse momentum phase-space of the self-injected electron bunch residing within the bubble. The B-field profile is schematically represented on the top of the figure. The laser driver moves from left to right as indicated by the arrow.}  
\end{center}
\end{figure}

To investigate controlled self-trapping in the presence of external static magnetic fields we present in this section 3D particle-in-cell simulations of laser (LWFA) and plasma (PWFA) wakefield accelerators using the particle-in-cell code Osiris. Figure~\ref{fig:lwfa} illustrates the evolution, and highlights the key mechanisms of injection assisted by external B-fields in the LWFA. The simulation window moves at the speed of light, with dimensions of $24\times24\times 12~(c/\omega_p)^3$, divided into $480\times480\times1200$ cells with $1\times1\times2$ electrons per cell in the $(x,y,z)$ directions respectively. The plasma ions are immobile. A linearly polarized laser pulse with central frequency $\omega_0/\omega_p=20$ was used, with a peak vector potential of $a_0=3$, a duration $\omega_p\tau_{\mathrm{FWHM}}=2\sqrt{a_0}$, and a transverse spot size matched to the pulse duration such that $W_0=c \tau_{\mathrm{FWHM}}$~\cite{bib:lu_prstab_2007}. The plasma density is of the form $n=n_0(z)\left(1+\Delta n r^2\right)$ for $r<\sqrt{10}~c/\omega_p$ and $n=0$ for $r>\sqrt{10}c/\omega_p$ with $\Delta n = \Delta n_c = 4/W_0^4$ (i.e. the normalized matching condition given by $\Delta n_c=4/(\pi r_e W_0^2)$, where $r_e=e^2/m_e c^2$ is the classical electron radius) being the linear guiding condition in the normalized units, and where $n_0(z)$ is a linear function of $z$ which increases from $n_0=0$ to $n_0=1$ for $50~c/\omega_p$ to ensure a smooth vacuum-plasma transition. The channel guides the front of the laser thereby minimizing the evolution of the bubble. A static external B-field pointing in the positive y-direction was used. At the point where the plasma density reaches its maximum value, the external field rises with $B_y^{\mathrm{ext}}=\omega_c/\omega_p = 0.6 \sin^2[\pi z / (2 L^{\mathrm{ramp}})+\Phi_1]$, with $L^{\mathrm{ramp}}=10 c/\omega_p$. It is constant and equal to $B_{y}^{\mathrm{ext}}=0.6$ for $L^{\mathrm{flat}}=40~c/\omega_p$ and then drops back to zero with $B_y^{\mathrm{ext}}=0.6 \sin^2[\pi z /(2 L^{\mathrm{ramp}})+\Phi_2]$. Moreover, $\Phi_1$ and $\Phi_2$ are phases chosen to guarantee the continuity of the external B-field profile.
Qualitatively, the longitudinal profile of the magnetic field thus resembles that of~\cite{bib:pollock_rsi_2006}.


Self-injection is absent from the regions where the B-field rises. In these regions, electrons traveling backwards ($v_z<0$) near the back of the bubble feel an increasing $\mathbf{v}\times \mathbf{B}^{\mathrm{ext}}$ force that rotates electrons anti-clockwise thereby locally decreasing (increasing) the blowout radius for $x>0$ ($x<0$). Then, as the B-field rises, the local wake phase velocity at the back of the bubble increases (decreases) for $x>0$ ($x<0$). For $x>0$, $v_{\phi}$ is superluminal, $\int \mathrm{d}_{\xi}\mathcal{H}\mathrm{d}\xi<0$, and self-injection can not occur. For $x<0$ trapping is precluded because electrons reach the axis in regions where the plasma focusing and accelerating fields are unable to focus and trap electrons. Thus, although for $x<0$ $\int \mathrm{d}_{\xi}\mathcal{H}\mathrm{d}\xi>0$, we have that $\int \mathrm{d}_{\xi}\mathcal{H}\mathrm{d}\xi+\Delta \psi^{\mathrm{ext}}<0$. 

Self-trapping occurs in the uniform regions of the external magnetic field where $x>0$. For $x>0$, electrons rotating anti-clockwise reach the axis in regions of maximum focusing and accelerating fields with larger $p_{\|}$ and can be trapped. For $x<0$, electrons reach regions of the axis (where focusing and accelerating fields are lower) with lower $p_{\|}$, and are are lost to the surrounding plasma. A threshold B-field for injection may be retrieved in the limit where $\gamma_{\phi}\rightarrow\infty$. Neglecting the plasma fields ($\Delta\psi^{\mathrm{pl}}=0$), and noting that the external longitudinal vector potential $A_{\|}^{\mathrm{ext}}=-B_y x$ is consistent with the considered magnetic field, leads to a simplified trapping condition $\Delta\psi^{\mathrm{ext}}= - B_y \Delta x = 1$, where $\Delta x=x_f-x_i\simeq -r_b$, where $r_b$ is the blowout radius and where it was considered that the initial (final) trapped electron radial position is $x=r_b$ ($x=0$). It shows that injection is facilitated in the region where $\Delta x<0$ is minimum. As a consequence, injection occurs off-axis (for $x>0$), and in a well defined azimuthal region defined by $- B_y r_b \sin\theta = 1$, where $\theta$ is the angle between the plane of the electron trajectory with the B field~\cite{bib:vieira_prl_2011}. Note, however, that this trapping threshold condition overestimates the threshold B-field for self-injection because it neglects the plasma fields. 

There is an upper $B_y$ value, given by $\omega_c/\omega_p\lesssim1$, beyond which injection may be suppressed in regions where the B field is flat. The later condition ensures that the plasma wakefields are nearly undisturbed by the external fields. Simulations then showed that when $\omega_c/\omega_p\gg 1$ there is a suppression of the wakefields that prevents injection. Hence trapping can be relaxed in the regions of uniform B-fields provided that $1/r_b \lesssim B_y\lesssim 1$ or, equivalently $170/r_b[10\mu\mathrm{m}]\lesssim B[\mathrm{T}] \lesssim 32\sqrt{n_0[10^{16}\mathrm{cm}^{-3}]}$.


The above-mentioned upper B-field limit for injection is absent from the downramp regions, where a stronger self-injection burst occurs for $x>0$. Injection occurs within the same angular and radial region as in the uniform B-field section (Fig.~\ref{fig:lwfa}c). For $x>0$, when the B-field lowers $r_b$ increases, $v_{\phi}$ lowers and $\int \mathrm{d}_{\xi}\mathcal{H}\mathrm{d}\xi>0$, facilitating injection. For $x<0$, $v_{\phi}>1$, and trapping is suppressed. The resulting phase space at $t=126 c/\omega_p$ is shown in Fig.~\ref{fig:lwfa}d.

After the magnetized plasma region, the magnetically injected electron bunch is clearly detached from the back of the bubble, leading to the generation of a quasi-mono-energetic electron bunch. The magnetic injected electron bunch right after the B-field is shown at $t=159 c/\omega_p$ in Fig.~\ref{fig:lwfa}e, and the corresponding phase-space in Fig.~\ref{fig:lwfa}f. The inset of Fig.~\ref{fig:lwfa}f shows the transverse phase space of the magnetically injected electron bunch residing within the blowout region. The asymmetrical distribution results from the fact that the injection process occurs off-axis. At this location, the emittance of the beam is on the order of 1$\pi$ mm~mrad in both transverse directions. Although comparison of beam emittance with a similar scenario without the B-field is not meaninfull because without the field the amount of self-injected charge is much smaller. However, the measured beam emittance is at the same values or lower than typical emittances of LWFAs.

\begin{figure}[htbp]
\begin{center}
\includegraphics[width=\columnwidth]{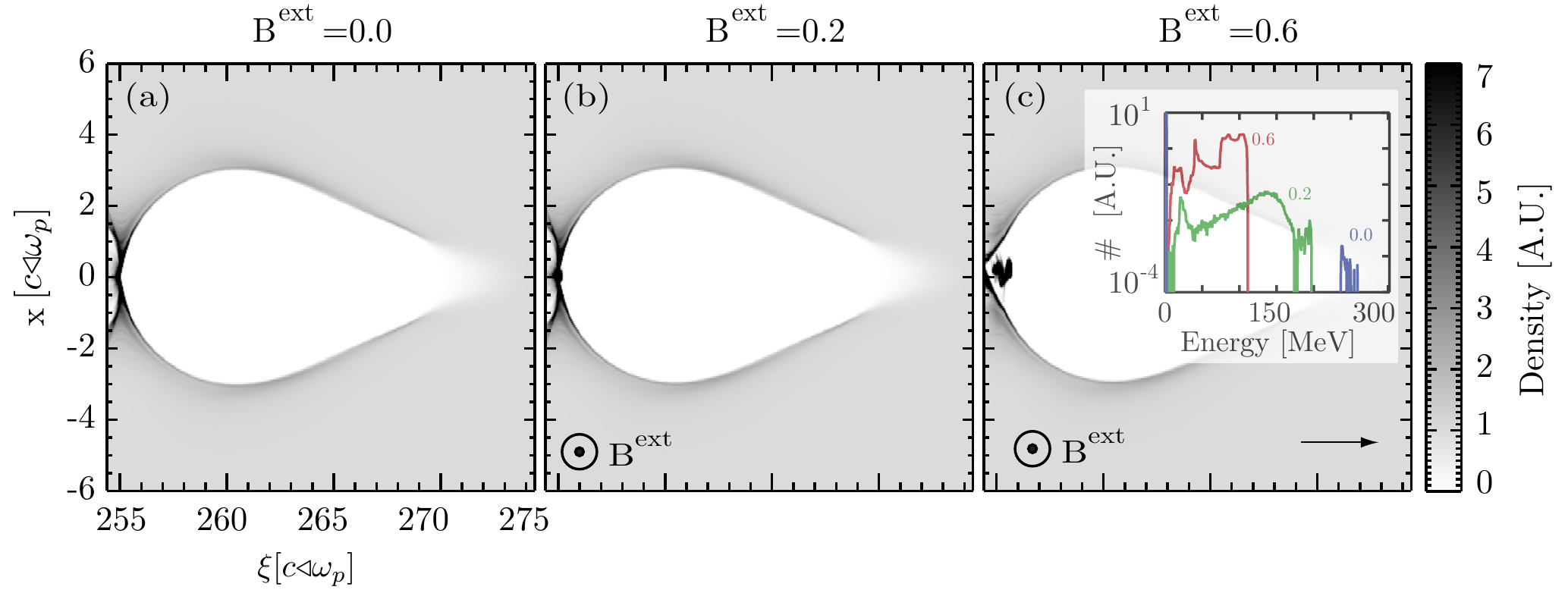}
\caption{\label{fig:pwfa} 3D osiris simulation results illustrating magnetic self-injection in the plasma wakefield accelerator using (a) $B^{\mathrm{ext}}=0$, (b) $B^{\mathrm{ext}}=0.2$, and (c) $B^{\mathrm{ext}}=0.6$ at $t=275/\omega_p$. In this scenario the B-field injection mechanism is dominated by the plasma bubble dynamics in the B-field down-ramp. It is clear that the trapped charge increases with the amplitude of the external field. The driver moves from left to right as indicated by the arrow in (c).}  
\end{center}
\end{figure}

External magnetic fields also relax the self-trapping thresholds in the PWFA. Figure~\ref{fig:pwfa} shows results from a 3D simulation of a magnetized PWFA. A 30 GeV electron bunch was considered with density profile given by $n_b=n_{b0}\exp\left(-\mathbf{x}_{\perp}^2/(2 \sigma_{\perp}^2)\right)\exp\left(-\xi^2/(2 \sigma_{\xi}^2)\right)$, with $\sigma_{\perp}=0.17~c/\omega_p$, $\sigma_{\xi}=1.95~c/\omega_p$, and $n_b/n_0=19$. These parameters ensure that $r_b$ is similar to the magnetized LWFA investigated above. The simulation window dimensions are $24\times24\times 24~(c/\omega_p)^3$, and it is divided in $480\times480\times640$ cells with $1\times1\times2$ electrons per cell in the $(x,y,z)$ directions respectively. The magnetic field profile is similar to the LWFA case.

The accelerating structures are similar for the LWFA and PWFA parameters described above since the blowout radius are similar for both cases. However, as shown by Eq.~(\ref{eq:btrapping}), self-injection thresholds are harder to meet in the PWFA than in the LWFA because $\gamma_{\phi}^{\mathrm{PWFA}}\simeq 60000 \gg \gamma_{\phi}^{\mathrm{LWFA}}\simeq 20$. In contrast to the LWFA scenario, injection is then absent in the PWFA in the uniform regions of the B-field, where the larger $p_z$ at the back of the bubble for $x>0$, associated with the additional electrons $\mathbf{v}\times\mathbf{B}$ anti-clockwise rotation, is still below that required for injection. Magnetic self-injection occurs only in the B-field downramp (Fig.~\ref{fig:pwfa}b-c), where the injection mechanism is similar to that ascribed to the LWFA. In general, stronger self-injection bursts occur in the B-field downramp for both LWFA and PWFA.

The amount of self-injected charge can be tuned by changing the B-field amplitude. The inset of Fig.~\ref{fig:pwfa}c shows the spectra of the self-injected charge in the first plasma bucket using $B^{\mathrm{ext}}=0.6$ (red curve), $B^{\mathrm{ext}}=0.2$ (green curve), and $B^{\mathrm{ext}}=0.0$ (blue curve). The amount of trapped charge is negligible in the un-magnetized scenario, and it is roughly 8 times larger for $B^{\mathrm{ext}}=0.6$ than for $B^{\mathrm{ext}}=0.2$ (notice that the plot is logarithmic in the vertical y-direction). These results show that higher B-field amplitudes increase the total amount of injected charge. 

Because of beam-loading~\cite{bib:tzoufras_prl_2008}, higher amounts of self-injected charge lead to lower accelerating gradients. Consequently, the maximum energy that can be achieved is lower for self-injected bunches with higher charges. This is consistent with the inset of Fig.~\ref{fig:pwfa}c which shows that self-injected bunches with lower charges reach higher energies. 

The inset of Fig.~\ref{fig:pwfa}c also shows that the energy spreads of the magnetically injected electrons are on the order of $100\%$. Due to the short duration of the self-injected bunch in comparison to the plasma wavelength, which guarantees uniform acceleration throughout the entire bunch length, the relative energy spread would decrease as the beam accelerates. Moreover, the energy spread would further narrow down near the dephasing length due to the bunch phase-space rotation~\cite{bib:tsung_prl_2004}.

For these parameters the threshold magnetic field for self-injection is $B_y^{\mathrm{ext}}\gtrsim0.2$. To connect these simulations with actual experimental conditions, we take $n_0=10^{17}~\mathrm{cm}^{-3}$ for which the electron beam and plasma parameters match those available at SLAC~\cite{bib:blumenfeld_nature_2007} with $\sigma_{\perp}=50.4~\mu$m, $\sigma_z=84~\mu$m and a total number of $3\times10^{10}$ electrons. For these parameters, $B_y^{\mathrm{ext}}=0.2$ corresponds to 20~T. These magnetic fields could be produced with state-of-the-art magnetic field generation techniques~\cite{bib:pollock_rsi_2006,bib:kumada_pac_2003}. By tuning further the plasma parameters controlled injection with magnetic fields as low as 5~T can also be achieved (cf. Sec.~\ref{sec:radiation}).



\section{\label{sec:tailor}Simultaneous generation of multiple self-injected electron bunches}

\begin{figure}[htbp]
\begin{center}
\includegraphics[width=\columnwidth]{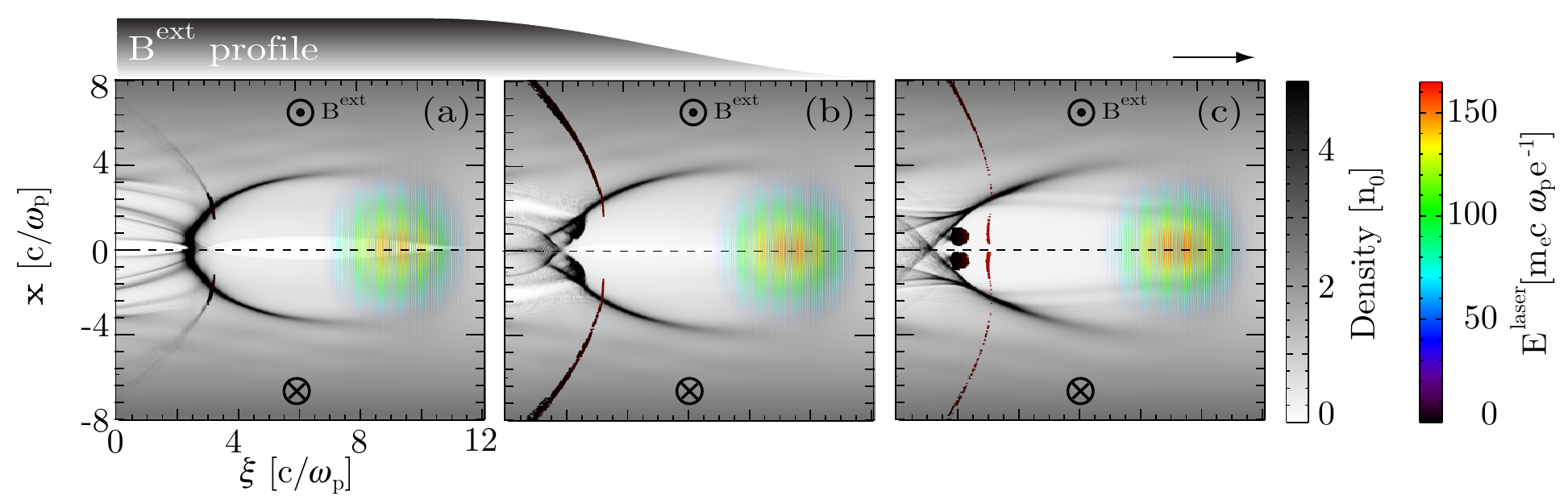}
\caption{\label{fig:tailor} 2D Osiris simulation of a magnetized LWFA using an azimuthal-like magnetic field. In the upper (lower) simulation mid-plane the B-field is directed outside (inside) the simulation plane. (a) is taken in the B-field flat region at $t = 79/\omega_p$, (b) in the B-field down-ramp at $t=92/\omega_p$, and (c) after the B-field at $t=102/\omega_p$. The B-field profile is schematically represented on the top of the figure. This configuration leads to identical magnetically injected electron bunches in the upper and lower simulation mid-plane. The laser driver moves from left to right as indicated by the arrow. The dark points represent self-injected particles.}  
\end{center}
\end{figure}

The transverse location where self-trapping is relaxed can be selected by adequate choice of the profile of the external magnetic field. As an example, Fig.~\ref{fig:tailor} shows the results from a 2D slab geometry simulation using 
a magnetic field which reverses sign at $x=0$  %
(this is equivalent to an azimuthal B-field profile in cylindrical symmetry).
In this case, the magnetic field points outside (inside) the simulation plane for $x>0$ ($x<0$). The 2D simulations use a simulation box that moves at $c$ with dimensions $12\times32~(c/\omega_p)^2$, and is divided into $640\times3000$ cells with $3\times3$ electrons per cell in the $(x,\xi)$ directions respectively. The laser pulse and plasma channel parameters are similar to those of the 3D LWFA simulation  (cf. Fig.~\ref{fig:lwfa}). The amplitude of the external B-field is $B_y^{\mathrm{ext}}=\omega_c/\omega_p = 0.6 \sin^2[\pi z / (2 L^{\mathrm{ramp}})+\Phi_1] x/|x|$, with $L^{\mathrm{ramp}}=10 c/\omega_p$, it is constant and equal to $B_{y}^{\mathrm{ext}}=0.6$ for $L^{\mathrm{flat}}=50~c/\omega_p$ and drops back to zero with $B_y^{\mathrm{ext}}=0.6 \sin^2[\pi z / (2 L^{\mathrm{ramp}})+\Phi_2] x/|x|$, where the choice of $\Phi_1$ and $\Phi_2$ ensures the continuity of the B-field longitudinal profile.

Figure~\ref{fig:tailor}a shows the magnetically injected electrons in the regions where the B-field is uniform. Two off-axis injection bursts occur at well defined transverse positions in the flat B-field regions. The two bunches are then injected symmetrically close to $x=0$. An additional and stronger self-injection burst occurs at the B-field down-ramp (Fig.~\ref{fig:tailor}b). After the magnetized plasma region, the two self-injected electron bunches continuously accelerate in the wakefield (Fig.~\ref{fig:tailor}c). Note that Fig.~\ref{fig:tailor}c refers to the early propagation of the electron bunch, much shorter than the dephasing length. Similarly, the propagation distance is much smaller than the betatron period of oscillation. The physical mechanisms under which self-injection occurs in the present configuration are identical to those presented in Sections~\ref{sec:review} and \ref{sec:simulations}. 

Interestingly, Fig.~\ref{fig:tailor} reveals that injection occurs in a highly spatially localized region. Off axis injection from well defined radial and azimuthal regions was observed in Sec.~\ref{sec:simulations} in 3D simulations. Generally, however, this effect is more noticeable in 2D slab geometry simulations than in 3D. These results also suggest that ring like electron bunches could be obtained in 3D. This could be advantageous for radiation generation purposes because bunch particles would perform betatron oscillations with similar amplitudes.


\section{\label{sec:radiation} Emission of betatron radiation at well defined frequencies}

Typical synchrotron radiation experiments in plasma accelerators reveal that radiation emission occurs in the wiggler regime. The wiggler regime enables emission of x-rays with broad spectra~\cite{bib:kneip_nphys_2012}. This contrasts with the undulator regime, where radiation is emitted at well defined harmonics. Although not yet attained experimentally, the undulator regime provides ideal conditions for radiation amplification, being critical for the realization of a ion-channel plasma based laser~\cite{bib:whittum_prl_1990}. This section illustrates how could magnetically injected electrons emit betatron radiation at well defined frequencies, closer to the undulator regime.

The PWFA beam and plasma simulation parameters, presented in Sec.~\ref{sec:simulations}, were tuned in order to lower the required magnetic field for injection, such that it could be more easily reached experimentally, and in order to lower the amplitude of the betatron oscillations in comparison to the plasma skin depth, such that distinguishable betatron radiation harmonics could be emitted. Systematic 3D parameter scans then showed that the threshold magnetic field for injection is 5.5 T at $n_0 = 10^{15}~\mathrm{cm^{-3}}$. At this plasma density, $L^{\mathrm{flat}}=40 c/\omega_p = 6.8$~mm, and $L^{\mathrm{ramp}}=10~c/\omega_p = 1.68$~mm, and the maximum B-field amplitude is $B_{y}^{\mathrm{ext}}=0.55~\omega_c/\omega_p$. These parameters are within current technological reach~\cite{bib:pollock_rsi_2006,bib:kumada_pac_2003}. Simulations used a simulation box with $12\times12\times16~(c/\omega_p)^3$, divided into $480\times480\times640$ cells with $2\times2\times1$ particles per cell for the electron beam and background plasma. 

\begin{figure}[htbp]
\begin{center}
\includegraphics[width=\columnwidth]{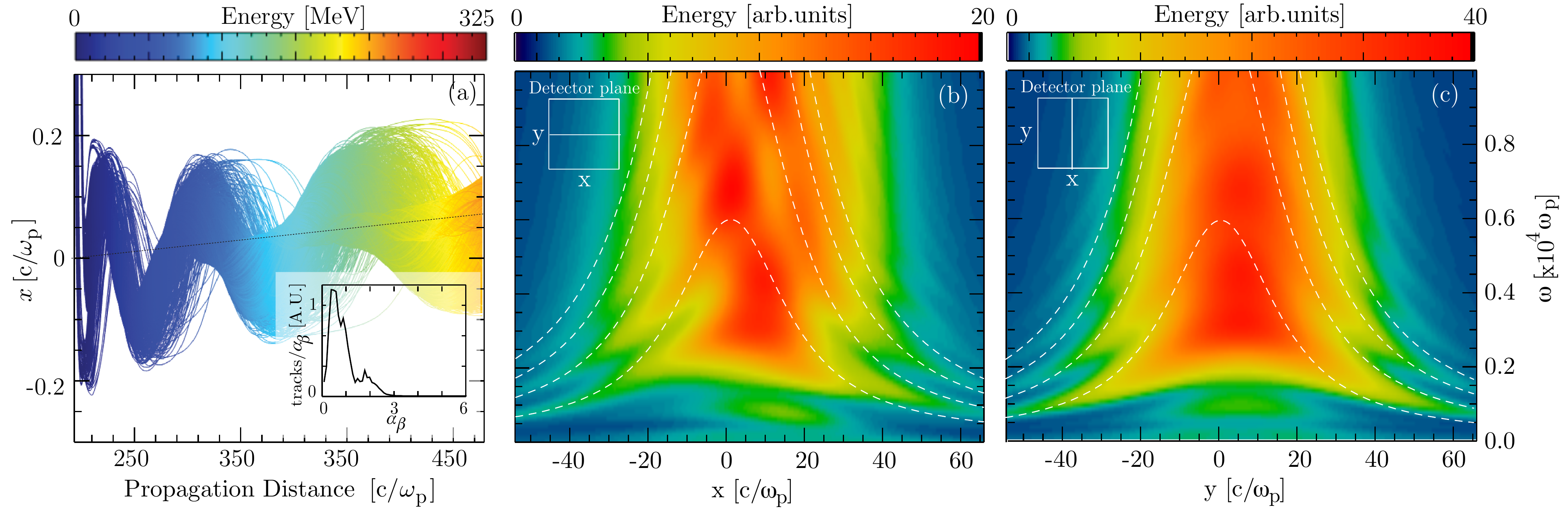}
\caption{\label{fig:tracks_pwfa} (a) B-injected electron bunch trajectories in the plasma wakefield accelerator. The trajectories gentle rise in the $x_2$ direction is due to the subtle transverse velocity shift imprinted on the particle beam driver in the B-field region. The inset represents the distribution of the radiation strength parameter $\alpha_{\beta}$. A significant fraction ($83\%$) of the trajectories are characterized by $\alpha_{\beta}\ll 1$. (b)-(c) Corresponding radiation profile lineouts passing through the center of a virtual detector located $5100 c/\omega_p$ from the end of the plasma. The white dashed lines corresponds to the predictions of Eq.~(\ref{eq:harmonics}) using $\gamma=400$ and $r_{\beta}=0.06$.The location where the spectra were taken in the detector plane is also shown.}
\end{center}
\end{figure}

Self-trapping occurs off-axis at the B-field downramp. Injection is localized radially and azimuthally, enabling the bunch to perform synchronized betatron oscillations (Fig.~\ref{fig:tracks_pwfa}a). The transverse x-axis shift of the electron trajectories result from the electron beam driver deflection when traversing the magnetized plasma region. The deflection angle is small and could be corrected by adding additional magnetized plasma regions with alternating B-fields along the propagation direction~\cite{bib:vieira_prl_2011}.

The small blowout radius ($r_b \simeq 1.5 c/\omega_p$) ensures that the betatron amplitudes of oscillation $r_{\beta}$ are much smaller than the plasma skin-depth ($\langle r_{\beta}\rangle \simeq 0.06 c/\omega_p$). The corresponding radiation strength parameter ($\alpha_{\beta}=\gamma K_{\beta} r_{\beta} k_p$) distribution, where $K_{\beta}=1/\sqrt{2 \gamma}$ is the normalized betatron frequency, and $k_p$ the plasma wavenumber, is shown in the inset of Fig.~\ref{fig:tracks_pwfa}a. It shows that a significant portion ($83\%$) of the electrons radiate with $\alpha_{\beta}<1$, an indication that single harmonics could be distinguishable in the emitted radiation spectrum. To retrieve the radiation spectrum, a random sample of the self-injected electrons was post-processed using the radiation code JRad~\cite{bib:martins_spie_2009}. Figure~\ref{fig:tracks_pwfa}b-c shows the radiation spectrum in the transverse central lines of a virtual detector placed at a distance $5100~c/\omega_p$ from the exit of the plasma. The detector lies on the $x-y$ plane.

The asymmetries in the x direction depicted in Fig.~\ref{fig:tracks_pwfa}b result from the tilt of the magnetically injected electron trajectories. Figure~\ref{fig:tracks_pwfa}b-c reveal that radiation is emitted at well defined frequencies, which are particularly clear at larger angles, i.e. for larger $|x|$. The width of each harmonic present in Fig.~\ref{fig:tracks_pwfa}b is larger than that expected in an idealized scenario, where radiation would be purely emitted in the undulator regime. This widening is due to the spread on the $\alpha_{\beta}$ distribution (through $\gamma$ and $r_{\beta}$ spreads) and also because some electrons radiate with strength parameters which are larger than unity $\alpha_{\beta}\gtrsim 1$. 


For an electron bunch with constant relativistic factor $\gamma$, and constant $r_{\beta}$ in a pure ion-channel, the frequency of the betatron radiation harmonics emitted in the undulator regime are given by~\cite{bib:esarey_pre_2002}: 
\begin{equation}
\label{eq:harmonics}
\frac{\omega_n}{\omega_p} = \frac{2 n \gamma^2 K_{\beta}}{\left(1+\alpha_{\beta}^2/2\right)\cos\theta + 2\gamma^2\left(1-\cos\theta\right)},
\end{equation} 
where $n$ corresponds to the $n^{\mathrm{th}}$ emitted harmonic, and $\theta$ to the angle between the velocity vector of the electron and the point in the detector. Radiation collected on-axis only exhibits odd-harmonics. To compare the predictions of Eq.~(\ref{eq:harmonics}) with simulation results we computed the particles trajectories average $\langle\gamma\rangle=400$ and $\langle r_{\beta}\rangle=0.06$. This yields $\alpha_{\beta}\simeq 0.7$, consistent with the inset of Fig.~\ref{fig:tracks_pwfa}. The analytical prediction Eq.~(\ref{eq:harmonics}) are shown by the dashed lines in Fig.~\ref{fig:tracks_pwfa}b-c. Eq.~(\ref{eq:harmonics}) is in good agreement with the simulation results specially for larger values of $|x|$. Discrepancies are due to the fact that the beam trajectories are tilted, and that $\alpha_{\beta}$, $r_{\beta}$, and $\gamma$ vary in time and for each electron.

\section{\label{sec:conclusions}Conclusions}

In conclusion, we explored further a recent controlled injection technique that uses transverse, static magnetic fields to tailor transverse properties of self-injection. This scheme leads to off-axis self-injection in well defined radial and azimuthal regions. A configuration consisting of a section of transversely uniform magnetized plasma yielding off-axis self-injection was investigated. It was shown that simultaneous self-injection of electron bunches could be achieved by using transversely non-uniform fields. This work also suggests that a series of magnetized regions could be used to produce a temporal sequence of electron bunches. Moreover, multiple spatially separated electrons could be produced simultaneously with transversely non-uniform B-fields. We showed that this technique could be used to produce electron bunches capable to emit betatron radiation at well defined frequencies with current technology.

\section{Acknowledgments}

Work partially supported by FCT (Portugal) through the grants SFRH/BPD/71166/2010, and PTDC/FIS/111720/2009, by the European community through LaserLab-Europe/Charpac EC FP7 Contract No. 228464. The simulations were performed at the IST Cluster, at Jaguar supercomputer under INCITE and on the JuGENE supercomputer.

\end{document}